\documentclass[a4paper]{article}
\usepackage{amsmath}
\usepackage{latexsym}
\usepackage{graphicx}
\usepackage{mathptmx}
\usepackage{bm}

\title{The Young-Laplace's equation for solid}

\author{Zaixing Huang\\[5pt]
  State Key Laboratory of Mechanics and Control of Mechanical Structures\\
  Nanjing University of Aeronautics and Astronautics\\
  Yudao Street 29, Nanjing, 210016, P R China\\ E-mail: huangzx@nuaa.edu.cn}

\begin{document}

\maketitle

\begin{abstract}
The Young-Laplace's equation is established based on liquid membrane
without shearing resistance. It is not valid for solid. By taking
into account the in-plane shearing and transverse shearing within
the surface layer, we reconstruct the Young-Laplace's equation so as
to characterize the surface of solid. A new version of the
Young-Laplace's equation is proposed. It shows that the surface
equilibrium of solid is determined by the bulk stress, surface
membrane stress and surface transverse stress together. The
transverse shear stress depends on the gradient of the Gaussian
curvature of surface and strain. The intrinsic membrane stress and
surface transverse shear stress cause the residual stresses to
appear in the interior of solid. The intrinsic surface transverse
shear stress only occurs in the non-spherical body.\\
\textbf{Key words:} Young-Laplace's equation, generalized
Young-Laplace's equation, surface stress, surface transverse stress,
Tolman' formula
\end{abstract}

%  The paper
\section{Introduction}
\label{intro} Surface is a thin layer with finite thickness rather
than a film of zero-thickness. For liquid, its surface can be
perfectly represented by a film only subjected to tension, because a
liquid can not support shearing stress indefinitely.
characterization to equilibrium of the liquid film leads to the
Young-Laplace's equation \cite{1}. On the other hand, a solid can be
in equilibrium under a shear stress. By introducing the in-plane
shear deformation, Gurtin and Murdoch extended the Young-Laplace's
equation into the generalized Young-Laplace's equation so as to
characterize the surface of elastic solid \cite{2}. Steigmann and
Ogden further proposed a reinforced generalized Young-Laplace
equation by taking into account the bending stiffness of the surface
film \cite{3}. Recently, Javili and Mosler et al revisited and
carefully examined the surface/interface elasticity theory \cite{4}.

So far, various models based on the Young-Laplace equationand the
generalized Young-Laplace equation have been presented in several
contexts. For example, Wang and Feng \cite{5} investigated the
influence of surface elasticity and residual surface tension on the
natural frequency of micro beams. It is not the purpose of this
short letter to list and review these abundant works. The reader can
refer to the reviews by Wang et al \cite{6}, Muller and Saul
\cite{7}, Sun \cite{8} and Duan et al \cite{9} on the relevant
literature.

However, all works mentioned above are developed based on the model
of film with zero-thickness. As a result, in the existing theories
and models it is impossible to take into account transverse shearing
effects within the surface layer of solid. In fact, since
non-uniformity of the excess energy profile across the surface layer
causes the energy gradient to appear, the transverse shear stress
inevitably exists on the cross-section of surface layer. Meanwhile,
if there is shearing on the internal boundary surface of the surface
layer, it will also cause the transverse shearing effects to occur
within the surface layer. To the best of our knowledge, hardly
anyone realizes existence of the transverse shear stress and its
influence on the equilibrium of a solid surface, while this
influence can not be neglected. Therefore, this problem will be
investigated in this paper, the emphasis will be placed on
reconstructing the Young-Laplace's equation for solid from the angle
of theory.

The paper is outlined as follows. In Section 2, we propose a
Lagrangian to describe the effects caused by the excess energy
within the surface layer and the Lagrangian equation and
curvature-dependent natural boundary condition. This boundary
condition is simplified into a new generalized Young-Laplace's
equation involved with the transverse shear stress in Section 3. In
Section 4, we analyze the characters of intrinsic membrane stress,
surface transverse stress and bulk residual stress and their mutual
relation. Finally, the summary and comment on the results in
this paper are given.\\

\textbf{Notation:} The index rules and summation convention are
adopted. Latin indices run from 1 to 3. The Greece letter $\Omega$
stands for a bounded domain of $R^3$, and $\partial\Omega$ is the
boundary surface of $\Omega$. The covariant derivative with respect
to coordinates is represented by the symbol $\partial_k$. The
contravariant derivative operator corresponding to $\partial_k$ is
denoted by $\partial^k=g^{kj}\partial_j$, where $g^{kj}$ is the
metric tensor. The symbol $\partial_A$ ($A=1, 2$) or $\nabla_s$ is
the surface gradient operator defined on $\partial\Omega$. The
derivative with respect to time is denoted by an upper dot, e.g.,
$\dot{a}={\textrm d}a/{\textrm d}t$. Other symbols will be
introduced in the text where they appear for the first time.

\section{Lagrangian field with surface effect}
\label{sec:1} Let $\textbf{x}=\{x^k\}$ be a 3-dimensional position
vector in $\Omega$ and $t\in[t_0, t_1]$ be time. A vector field
defined on $[t_0, t_1]\cup\Omega$ is denoted by $\phi_k=\phi_k(t,
\textbf{x})$. The Lagrangian of the field $\phi_k$ is written as
$L=L(\phi_k,\dot{\phi}_k,
\partial_j\phi_k)$.

Let spatial domain $\Omega$ occupied by $\phi_k$ be bounded and the
surface $\partial\Omega$ of $\Omega$ be a smooth surface. We believe
that physical behaviors of $\phi_k$ in the interior of $\Omega$ are
different from those on the boundary of $\Omega$. An additional
Lagrangian $\Gamma$ is used to characterize the physical behaviors
of $\phi_k$ on the boundary surface $\partial\Omega$. We refer to
$\Gamma$ as the surface Lagrangian, which is supposed to have the
form below
\begin{equation}\label{s1}\Gamma=\gamma(\phi_k,\dot\phi_k,\partial_A\phi_k)+
\nabla_s\cdot\mathbf{S}(\phi_k,\dot\phi_k,\partial_A\phi_k).\end{equation}   %1
On $\partial\Omega$, the vector
$\mathbf{S}(\phi_k,\dot\phi_k,\partial_A\phi_k)$ can be decomposed
into $\mathbf{S}(\phi_k,\dot\phi_k,\partial_A\phi_k)=
S^A(\phi_k,\dot\phi_k,\partial_A\phi_k)\mathbf{g}_A+\hat{\Gamma}(\phi_k,\dot\phi_k,\partial_A\phi_k)\mathbf{n}$,
where $\mathbf{g}_A$ is the unit base vector defined on the tangent
plane of $\partial\Omega$ and $\mathbf{n}$ the unit normal vector.
By the identity $\nabla_s\cdot\mathbf{n}=-2H$ \cite{10,11},
Eq.(\ref{s1}) is rewritten as
$\label{s2}\Gamma=\gamma+\partial_AS^A-2H\hat{\Gamma}$, where $H$ is
the mean curvature of $\partial\Omega$. Not losing generality, we
introduce a scale parameter $\chi(\textbf{x})$ which is defined as
the ratio of $2\hat{\Gamma}(\phi_k,\dot\phi_k,\partial_A\phi_k)$ to
$\gamma(\phi_k,\dot\phi_k,\partial_A\phi_k)$. As thus, we have
\begin{equation}\label{s2}\Gamma=(1-\chi H)\gamma+\partial_AS^A,\end{equation}        %2
By Eq.(\ref{s2}), the action of field can be represented as
\cite{12}
\begin{eqnarray}\label{s3}A[\phi_k]&=&\int_{t_0}^{t_1}\int_\Omega L(\phi_k,
\dot\phi_k,\partial_j\phi_k)\textrm{d}v\textrm{d}t+
\int_{t_0}^{t_1}\int_{\partial\Omega}\Gamma(\phi_k,\dot\phi_k,\partial_A\phi_k)
\textrm{d}a\textrm{d}t\nonumber\\
&=&\int_{t_0}^{t_1}\int_\Omega L\textrm{d}v\textrm{d}t+
\int_{t_0}^{t_1}\int_{\partial\Omega}[(1-\chi H)\gamma+\partial_AS^A]\textrm{d}a\textrm{d}t\nonumber\\
&=&\int_{t_0}^{t_1}\int_\Omega L\textrm{d}v\textrm{d}t+
\int_{t_0}^{t_1}\int_{\partial\Omega}(1-\chi H)\gamma\textrm{d}a\textrm{d}t,\end{eqnarray}          %3
where $\textrm{d}v$ and $\textrm{d}a$ are a volume measure in
$\Omega$ and an area measure on $\partial\Omega$, respectively. Let
$\delta\phi_k(t_0)=\delta\phi_k(t_1)=0$. Taking the variation of
$A[\phi_k]$ leads to
\begin{eqnarray}\label{s4}\delta A&=&\int_{t_0}^{t_1}\int_\Omega\{\frac{\partial L}{\partial
\phi_k}-\frac{\textrm{d}}{\textrm{d}t}\frac{\partial L}{\partial
(\partial\dot\phi_k)}-\partial_j[\frac{\partial L}{\partial
(\partial_j\phi_k)}]\}\delta\phi_k\textrm{d}v(x^k)\textrm{d}t\nonumber\\
&+&\int_{t_0}^{t_1}\int_{\partial\Omega}\{\frac{\partial
L}{\partial(\partial_j\phi_k)}n_j+(1-\chi H)(\frac{\partial
\gamma}{\partial\phi_k}-\frac{\textrm{d}}{\textrm{d}t}\frac{\partial
\gamma}{\partial\dot\phi_k}-\partial_A[\frac{\partial
\gamma}{\partial(\partial_A\phi_k)}])+\frac{\partial
\gamma}{\partial(\partial_A\phi_k)}\partial_A(\chi H)\}\delta\phi_k\textrm{d}a(x^k)\textrm{d}t,\end{eqnarray}      %4
where $n_k$ denotes the unit normal vector on $\partial\Omega$. The
Hamilton's principle asserts that $\delta A[\phi_k]=0$. Therefore,
according to the
fundamental lemma of variation, we have\\
Euler-Lagrange equation:
\begin{equation}\label{s5}\frac {\partial
L}{\partial \phi_k}-\frac{\textrm{d}}{\textrm{d}t}\frac{\partial
L}{\partial (\partial\dot\phi_k)}-\partial_j[\frac{\partial
L}{\partial(\partial_j\phi_k)}]=0,\quad x^k\in\Omega.\end{equation}                         %5
Natural boundary condition:
\begin{equation}\label{s6}\frac{\partial L}{\partial
(\partial_j\phi_k)}n_j=(1-\chi
H)(\frac{\textrm{d}}{\textrm{d}t}\frac{\partial
\gamma}{\partial\dot\phi_k}+\partial_A[\frac{\partial\gamma}{\partial
(\partial_A\phi_k)}]-\frac{\partial\gamma}{\partial\phi_k})-\frac{\partial
\gamma}{\partial(\partial_A\phi_k)}\partial_A(\chi H),\quad x^k\in\partial\Omega.\end{equation}             %6
Eq.(\ref{s5}) and (\ref{s6}) show that the surface Lagrangian has no
influence on the Euler-Lagrange equation, but it contributes to the
natural boundary condition and causes the natural boundary condition
to be correlated with the mean curvature and its gradient of
boundary surface. As a boundary condition, Eq.(\ref{s6}) is
universal but complicated. Next, we turn to simplification and
discussion to Eq.(\ref{s6}).

\section{The generalized Young-Laplace's equation and Tolman's length}
\label{sec:2}In the following, we stipulate that $\phi_k$ is a
displacement field. Only concerned with a quasi-static system,
Eq.(\ref{s6}) can be simplified into
\begin{equation}\label{p1}\frac{\partial L}{\partial
(\partial_j\phi_k)}n_j=(1-\chi
H)(\partial_A[\frac{\partial\gamma}{\partial
(\partial_A\phi_k)}]-\frac{\partial\gamma}{\partial\phi_k})-\frac{\partial
\gamma}{\partial(\partial_A\phi_k)}\partial_A(\chi H),\quad x^k\in\partial\Omega.\end{equation}                 %7
The surface Lagrangian $\gamma$ must be invariant under the
translational transformation of $\phi_k$. So $\gamma$ is necessarily
independent of $\phi_k$ itself, and Eq.(\ref{p1}) reduces to
\begin{equation}\label{p2}\frac{\partial L}{\partial
(\partial_j\phi_k)}n_j=\partial_A[(1-\chi
H)\frac{\partial\gamma}{\partial(\partial_A\phi_k)}],\quad x^k\in\partial\Omega.\end{equation}                  %8
Introduce two signs as follows
\begin{equation}\label{p3}\sigma^{kj}=\frac{\partial L}{\partial
(\partial_j\phi_k)},\quad \bar\sigma^{Ak}=\frac{\partial\gamma}{\partial(\partial_A\phi_k)}.\end{equation}      %9
In physics, $\sigma^{kj}$ and $\bar\sigma^{Ak}$ can be interpreted
as bulk stress and surface stress. By Eq.(\ref{p3}), Eq.(\ref{p2})
is represented as
\begin{equation}\label{p4}\sigma^{kj}n_j=\partial_A[(1-\chi
H)\bar\sigma^{Ak}],\quad x^k\in\partial\Omega.\end{equation}                  %10

Set a local coordinate system with the base vectors $(\mathbf{g}_1,
\mathbf{g}_2, \mathbf{g}_3)=(\mathbf{g}_A, \mathbf{n})$ on the
surface $\partial\Omega$, where $\mathbf{g}_A$ ($A=1,2$) is the the
covariant base vectors corresponding to the curvilinear coordinate
on $\partial\Omega$ and $\mathbf{n}$ the unit normal vector. In such
a coordinate system, we have
\begin{equation}\label{p5}\mathbf{\sigma}=\sigma^{kj}\mathbf{g}_k\otimes\mathbf{g}_j,\quad
\mathbf{\bar\sigma}_s=\bar\sigma^{Ak}\mathbf{g}_A\otimes\mathbf{g}_k=\bar\sigma^{AB}\mathbf{g}_A\otimes\mathbf{g}_B+
\bar\sigma^{A3}\mathbf{g}_A\otimes\mathbf{n}.\end{equation}                                       %11
Clearly, $\sigma^{AB}$ is the membrane stress component of surface
and $\sigma^{A3}$ is the transverse stress component on the
cross-section of surface layer. Let
\begin{equation}\label{p6}\mathbf{\sigma}_s=(1-\chi
H)\mathbf{\bar\sigma}_s,\quad x^k\in\partial\Omega.\end{equation}                                  %12
It is easy to see that Eq.(\ref{p6}) is just the Tolman's formula
\cite{13} in which $\mathbf{\bar\sigma}_s$ represents the surface
stress of a flat surface, while $\mathbf{\sigma}_s$ is the
curvature-dependent surface stress. So the scale parameter $\chi$ is
also referred as to the Tolman's length. The Tolman's formula has
been extensively applied to analyze the surface size effects of
micro/nano-scale liquid droplet and solid particle \cite{14,15}. By
Eq.(\ref{p5}) and (\ref{p6}), Eq.(\ref{p4}) can be equivalently
written as
\begin{equation}\label{p7}\mathbf{\sigma}\cdot\mathbf{n}=\nabla_s\cdot\mathbf{\sigma}_s,
\quad x^k\in\partial\Omega.\end{equation}                                                             %13
Eq.(\ref{p7}) is the so-called generalized Young-Laplace's equation,
but it is a new version taking into account the curvature effect and
transverse shearing effect of surface layer. To clarify this point,
firstly let us to calculate $\nabla_s\cdot\mathbf{\sigma}_s$ as
follows
\begin{equation}\label{p8}\nabla_s\cdot\mathbf{\sigma}_s=\{\partial_A[(1-\chi H)\bar\sigma^{AB}]-(1-\chi
H)\bar\sigma^{A3}b^B_A\}\mathbf{g}_B+\{\partial_A[(1-\chi
H)\bar\sigma^{A3}]+(1-\chi H)\bar\sigma^{AB}b_{AB}\}\mathbf{n},
\ x^k\in\partial\Omega,\end{equation}                                       %14
where
\begin{equation}\label{p8-1}\partial_A[(1-\chi H)\bar\sigma^{AB}]=[(1-\chi H)\bar\sigma^{AB}]_{,A}+(1-\chi
H)(\bar\sigma^{CB}\Gamma_{CA}^A+\bar\sigma^{AC}\Gamma_{AC}^B),\quad x^k\in\partial\Omega,\end{equation}            %15
\begin{equation}\label{p8-2}\partial_A[(1-\chi
H)\bar\sigma^{A3}]=[(1-\chi H)\bar\sigma^{A3}]_{,A}+(1-\chi
H)\bar\sigma^{A3}\Gamma_{AC}^C,\quad x^k\in\partial\Omega,\end{equation}                                       %16
where $\Gamma_{AC}^B$ and $b_{AB}$ (or $b^B_A$) are the connection
coefficients and curvature tensor of the surface $\partial\Omega$,
respectively. Substituting Eq.(\ref{p8}) into (\ref{p7}), and then
projecting it onto the tangential plane and normal direction of the
surface $\partial\Omega$, we have
\begin{equation}\label{p9}\mathbf{P}\cdot\mathbf{\sigma}\cdot\mathbf{n}=\{\partial_A[(1-\chi
H)\bar\sigma^{AB}]-(1-\chi H)\bar\sigma^{A3}b^B_A\}\mathbf{g}_B,
\quad x^k\in\partial\Omega,\end{equation}                                       %17
\begin{equation}\label{p10}\mathbf{n}\cdot\mathbf{\sigma}\cdot\mathbf{n}=\partial_A[(1-\chi H)\bar\sigma^{A3}]+(1-\chi
H)\bar\sigma^{AB}b_{AB},\qquad x^k\in\partial\Omega,\end{equation}         %18
where $\mathbf{P}$ is the projection operator, which reads
$\mathbf{P}=g^{ij}\mathbf{g}_i\otimes\mathbf{g}_j-\mathbf{n}\otimes\mathbf{n}$.
Eq.(\ref{p9}) and (\ref{p10}) are another form of the generalized
Young-Laplace's equation. It is obvious that they contain both the
surface transverse shearing effect and curvature-dependent effect of
membrane stress.

\section{Intrinsic membrane stress and surface transverse stress}
In a general case, although no external traction is prescribed, the
membrane stress $\bar\sigma^{AB}$ and surface transverse stress
$\bar\sigma^{A3}$ also exist due to the excess energy within the
surface layer. We refer to $\bar\sigma^{AB}$ and $\bar\sigma^{A3}$
as the intrinsic membrane stress and intrinsic surface transverse
stress if they are only caused by the excess energy within the
surface layer. For convenience, the intrinsic membrane stress and
intrinsic surface transverse stress are denoted by
$\bar\sigma^{AB}_0$ and $\tau^A$, respectively.

In terms of the Shuttleworth-Herring equation
$\bar\sigma^{AB}=\bar\gamma
g^{AB}+\partial\bar\gamma/\partial\phi_{AB}$, the intrinsic surface
transverse stress reads
\begin{equation}\label{p11}\bar\sigma^{AB}_0=\bar\sigma^{AB}|_{\partial_A\phi_B=0}=\bar\gamma g^{AB},\end{equation}  %19
where $\bar\gamma$ is the surface tension. Eq.(\ref{p11}) shows that
the intrinsic membrane stress always exists, irrelevant to the
curvature of surface. However, the intrinsic surface transverse
stress is different from the intrinsic membrane stress. Under some
special cases, the intrinsic surface transverse stress does not
occur. For example, no intrinsic surface transverse stress appears
on the surface of a spherical grain, due to the spherical symmetry.

Differential geometry tell us: a closed surface is a spherical
surface if and only if its Gaussian curvature is a constant
\cite{16}. It follows immediately that $\nabla_s\kappa=0$, where
$\kappa$ is the Gaussian curvature. As thus, the physical fact that
the intrinsic surface transverse stress does not on a spherical
surface but it occurs on a non-spherical surface shows that the
intrinsic surface transverse stress $\tau^A$ is related with
$\partial^A\kappa$. Meanwhile, $\tau^A$ is also dependent on the
shear modulus $\mu$. So under a general case, we have
$\tau^A=\partial^Af(\mu,\kappa)$. In terms of $\pi$ theorem of the
dimensional analysis \cite{17}, $f(\mu,\kappa)$ can be concretely
represented as $f(\mu,\kappa)=\mu\epsilon\kappa^{-1}$, where
$\epsilon$ is a dimensionless constant. Let $\tau=\mu\epsilon$.
Noticing that both $\mu$ and $\epsilon$ are constants, we have
\begin{equation}\label{p12}\tau^A=\bar\sigma^{A3}|_{\partial_A\phi_3=0}=\tau\partial^A\kappa^{-1}\end{equation}   %20
In physics, the constant $\epsilon$ can be interpreted as a
transverse shear strain caused by the excess energy within the
surface layer. Thus, $\tau$ is a residual shear stress on the cross
section of the surface layer. Substituting Eq.(\ref{p11}) and
(\ref{p12}) into (\ref{p9}) and (\ref{p10}) lead to
\begin{equation}\label{p13}\mathbf{P}\cdot\mathbf{\sigma}\cdot\mathbf{n}=\{\partial^B[\bar\gamma(1-\chi
H)]-\tau(1-\chi H)b^B_A\partial^A\kappa^{-1}\}\mathbf{g}_B,
\quad x^k\in\partial\Omega,\end{equation}                                       %21
\begin{equation}\label{p14}\mathbf{n}\cdot\mathbf{\sigma}\cdot\mathbf{n}=\partial_A[\tau(1-\chi H)\partial^A\kappa^{-1}]+
2\bar\gamma(1-\chi H)H,\qquad x^k\in\partial\Omega.\end{equation}         %22
Eq.(\ref{p13}) and (\ref{p14}) show that the excess energy within
surface layer can give rise to the residual stresses in the interior
of solid. It should be emphasized that the surface tension
$\bar\gamma$ differs from the surface Lagrangian $\gamma$. The
correlation between them can be represented as.
\begin{equation}\label{p15}\gamma=\int^{\partial_A\phi_B}_0(\bar\gamma g^{AB}+
\frac{\partial\bar\gamma}{\partial(\partial_A\phi_B)})\mathrm{d}(\partial_A\phi_B)+
\int^{\partial_A\phi_3}_0(\tau\partial^A\kappa^{-1}+
\frac{\partial\bar\gamma}{\partial(\partial_A\phi_3)})\mathrm{d}(\partial_A\phi_3).\end{equation}         %23
Inserting Eq.(\ref{p15}) in (\ref{p3})$_2$, we have
\begin{equation}\label{p16}\bar\sigma^{AB}=\frac{\partial\gamma}{\partial(\partial_A\phi_B)}=
\bar\gamma g^{AB}+\frac{\partial\bar\gamma}{\partial(\partial_A\phi_B)},\end{equation}                     %24
\begin{equation}\label{p17}\bar\sigma^{A3}=\frac{\partial\gamma}{\partial(\partial_A\phi_3)}=
\tau\partial^A\kappa^{-1}+\frac{\partial\bar\gamma}{\partial(\partial_A\phi_3)},\end{equation}                  %25
which are the constitutive equations characterizing the mechanical
behaviors of the solid surface.

For a liquid droplet in the static equilibrium, $\tau$ is identical
to zero and $\bar\gamma$ is a constant. Therefore, if $\chi=0$,
Eq.(\ref{p13}) and (\ref{p14}) reduce to
$\mathbf{n}\cdot\mathbf{\sigma}\cdot\mathbf{n}=2\bar\gamma H$. This
is just the original version of the Young-Laplace's equation for
liquid.

\section{Conclusion}
\label{sec:2}In the framework of the Lagrangian field theory, we
propose a surface Lagrangian to characterize the surface effects of
field, and reconstruct the generalized Young-Laplace's equation for
solid. Based on this equation, the conclusions are summarized as
follows.\\

1. On the surface of solid, there exists the transverse shear stress
induced by the excess energy within the surface layer. The
transverse shear stress depends on the gradient of the Gaussian
curvature of surface and deformation.

2. For the surface of a solid, its equilibrium is determined by the
bulk stress, surface membrane stress and surface transverse stress
together.

3. The intrinsic membrane stress and surface transverse shear stress
cause the residual stresses to appear in the interior of solid. The
intrinsic surface transverse shear stress only occurs in the
non-spherical body.\\

Finally, it should be pointed out that the influence of surface on a
bulk solid becomes obvious only at micro/nano scale. Meanwhile, it
also requires that the characteristic dimension of the solid must be
much larger than the thickness of surface layer so that the surface
layer can be treated as a surface of vanishing thickness. Otherwise,
atomistic or quantum models are necessary.

\section*{Acknowledgements}
The support of the National Nature Science Foundation of China
through the Grant No. 11172130 and 11672129 is gratefully
acknowledged.

\end{document}